\documentstyle[12pt,aaspp4,epsf,rotate]{article}
\newcommand{\illu}[4]
{\begin{figure}
\setlength{\epsfxsize}{#1mm}
\epsfverbosetrue
\vspace*{#2mm}
\epsffile{#3}
\caption{#4}\end{figure}}

\newcommand{\be }{\begin{equation}}
\newcommand{\kms}{km\,s$^{\rm -1}$}
\newcommand{\ee }{\end{equation}}

\newcommand{\ls}{L$_\ast$}
\newcommand{\lsun}{$L_{\odot}$}

\newcommand{\degr}{$^{\circ}$}

\newcommand{\SII}{[SII]$\lambda\lambda$6716,6731}

\newcommand{\HT}{H$_{\rm 2}$}

\lefthead{Micono et al.}
\righthead{Proper motions and variability in HH\,46/47}
\begin{document}

\title{PROPER MOTIONS AND VARIABILITY OF THE H$_2$ EMISSION IN THE HH~46/47
SYSTEM}
\author{Milena Micono$^{1,2}$, Christopher J.\ Davis$^{1,3}$,
Thomas P.\ Ray$^{1}$, Jochen Eisl\"offel$^{4}$ \& Matthew D.\ Shetrone$^{5}$}
\vspace*{0.5in}
\affil{$^1$School of Cosmic Physics, Dublin Inst.\ for Adv.\ Studies,
5 Merrion Square, Dublin 2, Ireland}
\affil{$^2$Dipartimento di Fisica Generale dell'Universit\`a, Via Pietro Giuria
1,
I-10125 Torino, Italy}
\affil{$^3$Joint Astronomy Centre, 660 N.\ A'oh\=ok\=u Place,  University
Park, Hilo, Hawaii 96720, U.S.A.}
\affil{$^4$Th\"uringer Landessternwarte Tautenburg,
Sternwarte 5, D-07778 Tautenburg, Germany}
\affil{$^5$ European Southern Observatory, Alonso de Cordova 3107, P.O.\ 19001,
Santiago 19, Chile}
\begin{abstract}
We report here on the first proper motion measurements of molecular hydrogen 
emission features in the Herbig-Haro 46/47 outflow.  Assuming a
distance of 350 pc to this flow, the inferred tangential velocities
range from a few tens to almost 500~\kms . The highest velocities are
observed for \HT\ knots either in, or close to, the jet/counterjet axes.
Knots constituting the wings of the large scale \HT\ bow (see, for example,
Eisl\"offel et al.\ 1994) are found to move much more slowly. These
results appear to be in agreement with recent numerical simulations
of \HT\ emission from pulsed jets. We also report the first detection of
variability in \HT\ features for a young stellar object (YSO) outflow. It was
found that several \HT\ knots significantly changed their luminosity over the
4 year timebase used to conduct our study. This is in line with current
estimates for the cooling time of gas radiating shocked \HT\ emission in YSO
environments.
\end{abstract}


\vspace*{1.0in}
\received{}
\accepted{}

\section{Introduction}
Although low mass star formation is accompanied by a diverse range of outflow
phenomena evident at many wavelengths, it seems likely that the primary driving
flow in each case is a Herbig-Haro (HH) jet (see, e.g., Padman, Bence
\& Richer 1997). Such jets were first discovered by Mundt \&
Fried (1983) and for a general review of their properties the reader
is referred to Edwards, Ray \& Mundt (1993). In regions of low
extinction, HH jets can be observed optically in a number of emission
lines, e.g.\ the \SII\ doublet.  In those cases were the jets are more
deeply embedded, however, near-infrared \HT\ emission is a much more
useful tracer of HH flows (Eisl\"offel 1997) as evidenced, for
example, by the dramatic \HT\ v=1--0 S(1) 2.12$\mu$m images of Cep~E
(Eisl\"offel et al.\ 1996) and HH~212 (Zinnecker et al.\ 1997).

The general consensus is that the \HT\ emission seen in young stellar
object (YSO) outflows is due to radiative shocks. 
There has been, however, considerable debate in the literature over the
nature of the shocks themselves with various observers preferring jump
(J-)type and others magnetic continuous (C-)type shocks to explain
their data (Eisl\"offel 1997). In particular J-shocks seem to be
supported by excitation analysis (Gredel 1996; Everett 1997) whereas
C-shocks are needed to model the observed wide (50--100 \kms, FWHM)
line profiles (e.g., Davis \& Smith 1996). At the same time, an
alternative scenario has been proposed by Taylor \& Raga (1995) in
which the \HT\ emission is due to collisional excitement in turbulent
boundary layers between the jet and its ambient medium. In such layers
one expects to find flow velocities ranging from that of the ambient
medium up to the velocity of the jet itself (Taylor \& Raga 1995;
Noriega-Crespo et al.\ 1997). Obviously in order to help distinguish
between these different models, full kinematic data is required for
the emitting regions. The combination of high resolution spectroscopic
observations coupled with imaging at a number of epochs to determine
proper motions and cooling timescales provides powerful constraints
that can aid in selecting the correct model. With this in mind we have
monitored, for a number of years, several well-known YSO outflows that
are visible in the 2.12$\mu$m \HT\ line. In order to obtain proper
motion data, one requires IR camera images that contain sufficient
reference stars. Such images have only been attainable in the past few
years, so only now is it possible to make the first proper motion and,
coincidentally, variability measurements of \HT\ features.

The outflow we report on here, the HH~46/47 system (Dopita, Schwartz
\& Evans 1982) has as its source the low luminosity \ls\ $\approx$
10\lsun\ young stellar object (YSO) IRAS 08242-5050. This outflow is
at the edge of a Bok globule, and powers not only a set of well-known
optically-visible HH objects (e.g., Eisl\"offel \& Mundt 1994) but
also a bipolar molecular outflow (Olberg, Reipurth \& Booth
1992). The northeastern (blueshifted) portion of the flow is best
observed optically (see, for example, the dramatic HST images in
Heathcote et al.\ 1996). To study, however, the embedded
``counterflow'' as it ploughs into the Bok globule, one has to use
infrared imaging.  Such imaging, in the 2.12$\mu$m \HT\ line, not only
shows quite strikingly the ``counterjet'' but also an extensive bow
shock that is clearly driven by the counterjet (Eisl\"offel et al.\
1994). 

The earliest detections of proper motions in HH~46/47 were made by Schwartz, 
Jones \& Sirk (1984). More recently a comprehensive optical proper 
motion study of this region was carried out by Eisl\"offel \& Mundt (1994). 
Here we report the first such study in \HT\ emission, and one of the first 
measurements of proper motion of \HT\ features in any YSO flow (see also
Noriega-Crespo et al.\ 1997). Variability in \HT\ emission is also
detected here for the first time.

\section{Observations and data reduction}

Data obtained at 3 epochs (April 1993, November 1995 and April 1997),
spanning a period of 4.0\,yrs, were used to calculate proper motions
in HH\,46/47.  These data have all been obtained at the ESO/MPI 2.2\,m
Telescope, using the IRAC2 near-IR camera which is equipped with a
256$\times$256 pixel NICMOS\,3 array and optics that yield a
0\farcs 52/pixel scale.  On each occasion, an H$_2$ v=1-0 filter
($\lambda = 2.121\,\mu$m, $\Delta\lambda = 0.04\mu$m) was employed.
The same standard data reduction (sky-subtraction and dome
flat-fielding) was carried out on each data-set; the observing
sequence and data reduction are described in Eisl\"offel et
al. (1994).

The reduced images were registered by the application of linear
geometric transformations (translation, rotation and re-scaling)
derived by comparing the positions of a number of field stars at
different epochs. The positions of the stars were matched with an
accuracy of the order of a tenth of a pixel.  To measure the positions
of the structures two methods were employed.  For small and well
defined knots, such as the jet/counterjet knots and similar bright
substructures in more extended faint regions, we compared their
positions at different times calculated using the intensity weighted
first moments of the pixel values; the errors arising from this
procedure were generally smaller than the error due to frame
alignment.  For more diffuse features, we calculated the shifts using
a cross correlation technique, similar to that employed by Reipurth \&
Heathcote (1991), between boxes drawn around each structure at the
different epochs. The errors associated with this procedure were
estimated as typically 0.1 pixels ($\sim$ 0\farcs 05)
for the brighter structures and 0.2 pixels ($\sim$ 0\farcs 1) for 
the fainter ones.

\section{Results} %%%%%%%%%%%%%%%%%%%%%%%%%%%%%%%%%%%%%%%%%%%%%%%%%

The proper motions of the  HH\,46/47 H$_2$ knots, calculated using a linear
least squares fit to the data points, are given in Table 1. They are also
illustrated in Figs.\,1 and 2, where we display, in contour form, an H$_2$ 
(+continuum) narrow-band mosaic of the HH\,46/47 field from the April
1997 run plus separate finer scale images of the H$_2$ counterjet.
We adopt a value of 350\,pc for the distance to HH~46/47
in agreement with Eisl\"offel \& Mundt (1994) and in line with
more recent estimates of the distance to the nearby Gum Nebula (Franco 1990).
In the direction of the counterflow, large tangential proper motions,
in excess of 200\,km~s$^{-1}$, are measured for the knot at the apex of the
bow (labeled a in Fig.\ 1 and lying in the vicinity of the optical knot
HH\,47C), the two knots within the bow, g and h, and the bright counter-jet
knot, c-jet-d (see Fig.\ 2). Much lower proper motions ($<$ 100\,km~s$^{-1}$)
are measured along the bow wings (knots b~--~f ). The H$_2$ proper motion
measurements imply flow speeds along the counter-jet in the range
170--280\,km~s$^{\rm -1}$, assuming its inclination angle with respect to
the plane of the sky is $\sim$ 35\degr\ as suggested by the optical data 
(Eisl\"offel \& Mundt 1994). A comparison of the tangential velocities of the 
\HT\ and optical knots in the counterflow shows that while both sets of 
features are moving in almost the same direction there is a tendency for the 
\HT\ knots to be moving somewhat faster. For example knot~a has a tangential 
velocity of 260\,km~s$^{-1}$ whereas the corresponding average velocity for 
the optical knots constituting HH\,47C is approximately 150\,km~s$^{-1}$ 
(Eisl\"offel \& Mundt 1994).

Turning to the northeastern section of the flow, this is far more prominent
in optical forbidden line emission, e.g.\ \SII , than in \HT\ emission.
Close to the optical jet, however, that links the source and HH\,46
with the HH\,47A bow shock, there is a faint H$_2$ emission filament
(x in Fig.\,1) as well as an ``arc'' of H$_2$ emission towards HH\,47A itself.
Davis, Eisl\"offel \& Smith (1996) associate the ``arc'' with the
wings of the HH\,47A bow shock (based on a comparison of its
morphology with synthetic H$_2$ images of C-type bow shocks). The
brightest peak in this H$_2$ arc, knot z, has a tangential velocity 
(67\degr , 184\,km~s$^{-1}$) that is very similar to its optical 
counterpart HH\,47A0 (45\degr , 182\,km~s$^{-1}$) as measured by Eisl\"offel 
\& Mundt (1994).  The fainter H$_2$ filament, x, has a somewhat lower velocity,
and interestingly seems to be moving almost northward, perhaps because
of the lateral expansion of the outflow lobe itself (Fig.\,1). Although there
does not appear to be any corresponding optical feature in the same region, 
nevertheless there are a number of optical knots (A6 and A7) to the south of 
HH\,47A which may also be expanding laterally (Eisl\"offel \& Mundt 1994).

When comparing our \HT\ images for the 3 epochs, it was noted that a
small number of knots either appeared for the first time or faded from
view within the 4 year period. Such knots were not used for proper motion
measurements. An example can be seen in Fig.\ 2 where one of the knots in the
counterjet, c-jet e, was seen both in the 1993 and 1995 images but
disappeared by 1997. The fading and brightening of knots can, we believe, 
also cause anomalous proper motion results. 
For example the measured proper motion for knot
c-jet b is not in the same direction as the counterjet and out of line
with  the motion of knots c-jet a and d. A cursory examination of Fig.\ 2,
however, shows that in Nov.\ '95 knot c-jet b probably consisted of two
marginally resolved sub-condensations, and that the most southwesterly one
faded by Apr.\ '97. The net effect is, of course, to rotate the apparent proper
motion vector more towards the source, as observed.

\section{Analysis and Discussion} %%%%%%%%%%%%%%%%%%%%%%%%%%%%%%%%%%%%%%%%%

It is well known that molecular hydrogen dissociates in planar J and
C-type shocks at velocities typically greater than $\sim$25 and
$\sim$50 \kms\ respectively (Smith 1994).  Given the large outflow
speeds inferred by our proper motion measurements, and assuming the
emission is from shocks, it is clear that the shocks are either
oblique, e.g.\ as in the wings of a bow shock, and/or that they are
caused by marginally faster material moving into somewhat slower
gas. If the latter case applies, the shock speed can be quite low even
though the velocity of the post-shock gas may be high. Numerical
simulations of the expected \HT\ emission from YSO outflows are only
now becoming available (Smith 1991; Suttner et al. 1997) and in
particular a simulation appropriate to the conditions for the
HH\,46/47 counterflow has recently been made (Downes \& Ray 1998).
Broadly speaking these simulations are in line with our
observations. For example they predict low tangential velocities for
the \HT\ knots in the bow shock wings as is observed. The bow emission
derives from entrained shocked ambient material and so, providing the
pre-shock material is virtually stationary, we would not expect its
post-shock velocity to be very high.  On the other hand the \HT\
emission within the flow itself can have much higher tangential
velocities if it arises from marginally higher velocity material
catching up with previously ejected gas. In this way, and
providing there is sufficient
\HT\ present, the \HT\ knots can reach tangential velocities close to that
of the jet itself, i.e.\ several hundred \kms . The high tangential
velocity of the \HT\ emission close to apex of the bow (knot~a) may
arise in a slightly different way as it could come from the Mach disk 
(the shock ``facing'' the outflow source that decelerates the jet gas). 
Overlaying our \HT\ plots onto the \SII\ images of Eisl\"offel \& Mundt 
(1994) shows that the \HT\ emission is located just {\em behind} the \SII\ 
bow as is consistent with this hypothesis.  Moreover, if the jet is moving into
lower density material near HH\,47C, the Mach disk could be a low
velocity shock.  Under these conditions we might expect the excitation, rather 
than the dissociation, of molecular hydrogen. In this regard, it is interesting 
to note that for the corresponding feature in the northeastern flow, i.e.\ the 
Mach disk associated with HH\,47A, Heathcote et al.\ (1996) estimate its shock 
velocity to be about 35 \kms .  Although this is close to the limit for the 
dissociation of \HT , if we are dealing with a J-type shock, the conditions 
near HH\, 47C might be somewhat more favorable for the molecule's survival. 

In all of the observed knots and filaments, the cooling time for the
heated \HT\ molecules is expected to be short ($\le 2$ years, assuming
$n_{\rm H2} \ge 10^{6}$\,cm$^{-3}$ and $T_{\rm k} \sim 2000$\,K; see
for example Davis \& Smith 1995). We therefore expect and, as we have
already mentioned, indeed find that a number of H$_2$
knots appear and disappear on such timescales, and that the
morphologies of some other knots change.  One should also realize that,
because the 4\,year time span of our observations likely exceeds the
cooling time, then our proper-motion measurements record the
tangential motions of the shocks themselves rather than of the post-shock 
gas, since the shock-heated gas observed in our epoch-1993
observations will, by 1997, have radiatively cooled and will therefore
no-longer be observable.  Moreover, if in this time a shock front encounters 
lower-density, or perhaps even purely atomic gas, then emission observed in a 
particular region in 1993 could seemingly ``disappear'' by 1997.  
Indeed, the speeds, directions and apparent morphologies of the
shocks themselves, as well as their ability to generate molecular line
emission, are strongly depending on the conditions in the pre-shock
gas, which may be inhomogeneous on small (even sub-arcsecond) scales.

\section{Conclusions}
Near-IR images, obtained over a 4\,year period with the IRAC\,2
common-user instrument at ESO, Chile, were used to compute the proper
motions of numerous H$_2$ knots and filaments in the HH\,46/47 bipolar
outflow. Tangential velocities ranging from 100--500\,km\,s$^{-1}$ were
recorded for the HH\,47A bow shock and for knots in, or close to, the IR 
counter-jet. Lower velocities were measured along the bow shock wings in the
counter-flow, in support of earlier interpretations of the overall flow
morphology and recent numerical simulations of molecular jets. The measured
tangential velocities of the \HT\ features appear comparable in magnitude, 
and generally follow the trend of the corresponding optical features in the 
same regions. Variability is also observed in some knots, which change 
apparent ``shape'' or vanish altogether.  This is a result of the fact that the
cooling times associated with the observed molecular shock features are
comparable to or shorter than the 4\,year time-span of our observations.

\bigskip
\noindent
{\bf ACKNOWLEDGEMENTS}

\noindent
We would like to thank the 2.2\,m team at ESO for their continued support
with the use of IRAC\,2.  M.M. wishes to acknowledge the warm
hospitality and financial support from DIAS where this work was carried out. 
Finally we would also like to thank the referee, Alex Raga, for his helpful
comments.\\

\newpage %%%%%%%%%%%%%%%%%%%%%%%%%%%%%%%%%%%%%%%%%%%%%%%%%%%%%%%%%%%%

\begin{table}[hp]
\begin{center}
TABLE 1 \\
{\sc Proper motions of the H$_2$ knots in HH\,46/47} \\

\begin{tabular}[t]{lccc} \hline \hline
 \HT\ Knot    & Shift    &  PA         &  $V_{\rm tan}^1$   \\
         & (arcsec/yr) & (degrees)   &   (km s$^{-1}$)    \\
\hline \\
 a            & 0.157($\pm$0.011)  &  254.9($\pm$3.8)   &  260  \\
 b            & 0.049($\pm$0.011)  &  252.8($\pm$12.3)  &   81  \\
 c$^2$        & 0.036($\pm$0.011)  &  256.0($\pm$16.5)  &   60  \\
 d            & 0.047($\pm$0.004)  &  311.8($\pm$5.1)   &   78  \\
 e            & 0.029($\pm$0.004)  &  217.1($\pm$8.2)   &   48  \\
 d+e$^2$      & 0.031($\pm$0.011)  &  285.5($\pm$19.6)  &   51  \\
 f            & 0.024($\pm$0.012)  &  201.9($\pm$26.1)  &   40  \\
 g            & 0.250($\pm$0.021)  &  266.7($\pm$4.8)   &  414  \\
 h            & 0.296($\pm$0.021)  &  254.0($\pm$4.1)   &  492  \\
 c-jet~a      & 0.082($\pm$0.019)  &  215.9($\pm$13.2)  &  136   \\
 c-jet~b$^3$  & 0.083($\pm$0.016)?  &  128.6($\pm$10.9)?  & 138?  \\
 c-jet~d      & 0.138($\pm$0.016)  &  237.9($\pm$6.6)   &  229  \\
 x            & 0.049($\pm$0.021)  &    343.6($\pm$24.4) &  81 \\
 y            & 0.189($\pm$0.005)  &     59.5($\pm$1.6) &  314 \\
 z            & 0.111($\pm$0.005)  &     66.6($\pm$2.7) &  184 \\
\\ \hline
\end{tabular}
\end{center}
$^1$Assuming a distance of 350\,pc to HH\,46/47.\\
$^2$Apparent motion of complex using the cross correlation
technique (see text). \\
$^3$For a possible explanation of these rather anomalous results see text \\
\end{table}

\newpage
\noindent
{\bf Figures: }

\figcaption[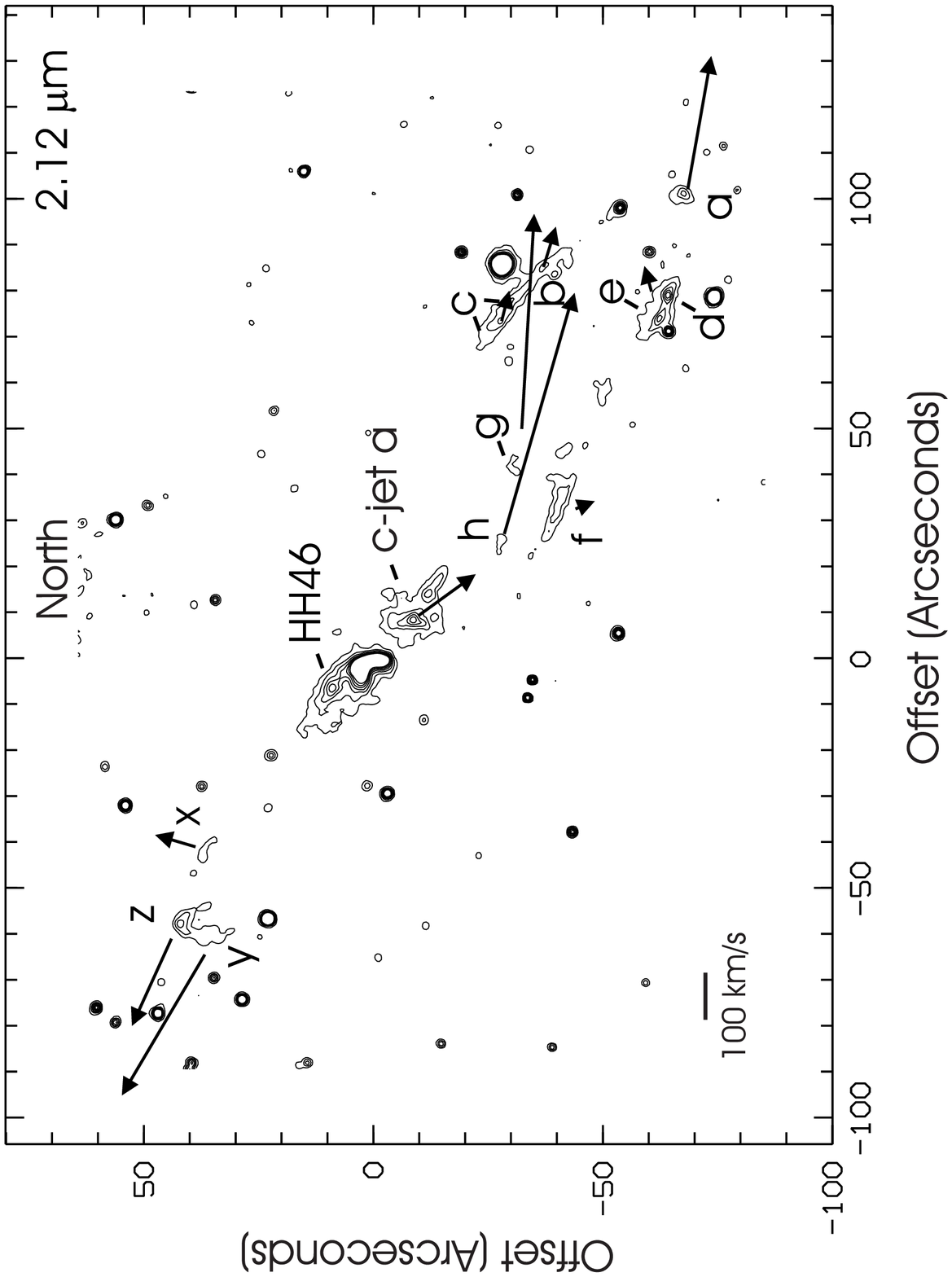]
{2.12$\mu$m wide-field contour plot of the HH\,46/47 outflow derived
from our image taken in April 1997 (see text). The various knots used
for the proper motion study are identified (see Table 1). The contour
levels are at 2, 4, 6, 8, ... 10$^{\rm -18}$W\,m$^{\rm -2}$arcsec$^{-2}$.
Proper motion vectors are shown. HH\,46, which is in part a reflection 
nebula, is indicated.}

\figcaption[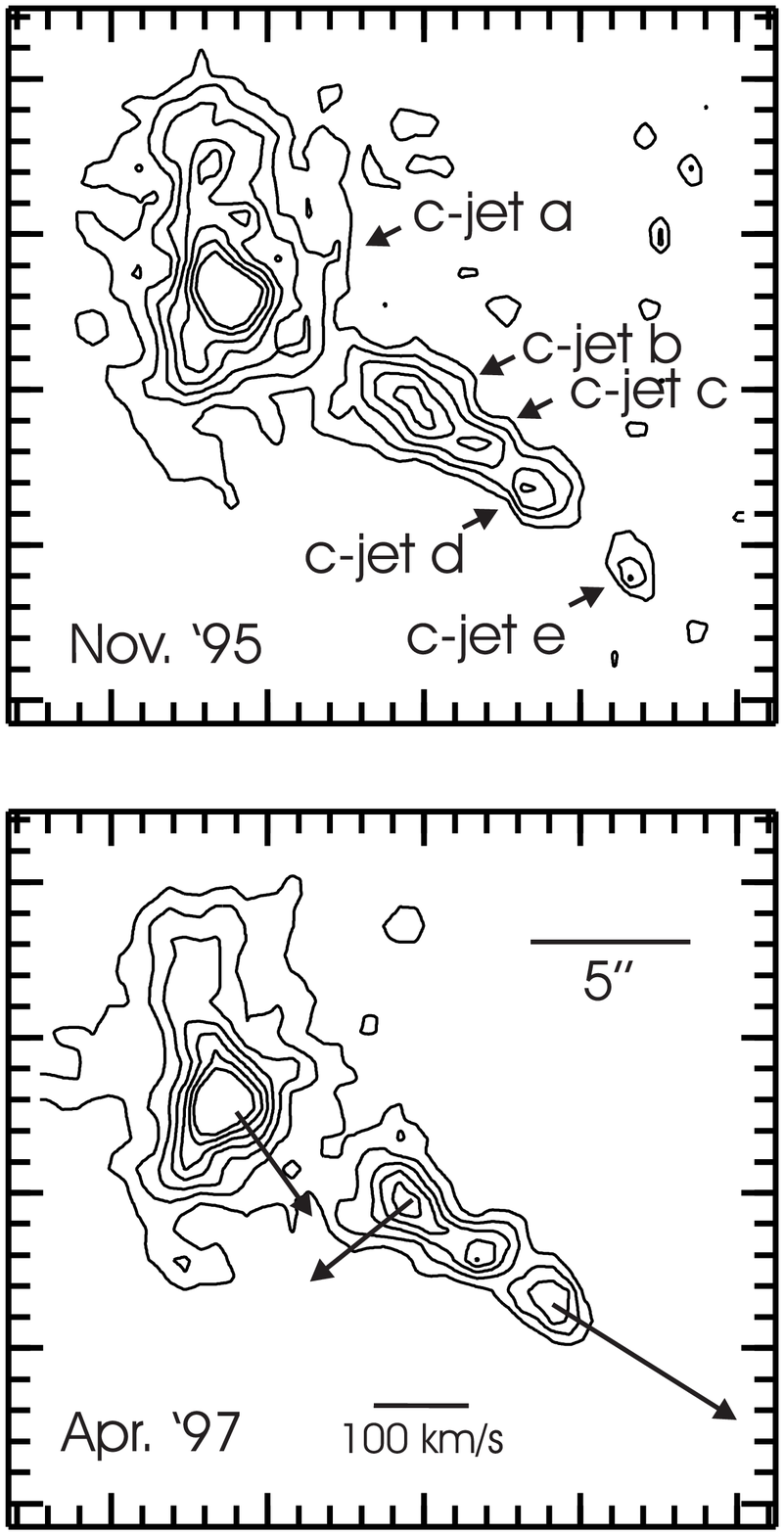]
{2.12$\mu$m contour plot of the HH\,46/47 counterjet at
two epochs (November 1995 and April 1997). As in Fig.\ 1 the knots referred to
in the text are identified and proper motion vectors are shown. Note the
disappearance of knot c-jet~e in the 1997 image. The somewhat spurious result
for knot c-jet~b is explained in the text. The contour levels increase in steps
of 1.2 from a base level of 2.4 $\times$ 10$^{\rm -18}$W\,m$^{\rm -2}$
arcsec$^{-2}$.}

\begin{figure}  Fig.\ 1
\begin{center}
\leavevmode
\epsfxsize=14.0cm
\epsfbox{fig1.eps}
\end{center}
\end{figure}

\begin{figure}  Fig.\ 2
\begin{center}
\leavevmode
\hspace*{-3.5cm}
\epsfxsize=14.0cm
\epsfbox{fig2.eps}
\end{center}
\end{figure}

\begin{references}
\reference{} Davis, C.J.,  Eisl\"offel, J., \& Smith M.D. 1996, \apj, 463,
246
\reference{} Davis, C.J., \&  Smith M.D. 1995, \aap, 443, L41
\reference{} Davis, C.J., \&  Smith M.D. 1996, \aap, 309, 929
\reference {} Dopita, M.A., Schwartz, R.D., \& Evans, I. 1982, \apj, 263,
L73
\reference{} Downes, T.P., \& Ray, T.P. 1998, in preparation
\reference{} Edwards, S., Ray, T.\ P., \& Mundt, R. 1993, in Protostars and
Planets III, eds.\ E.\ Levy \& J.\ Lunine, (University of Arizona Press), 567
\reference{} Eisl\"offel, J. 1997, in Herbig-Haro
Outflows and the Birth of Low Mass Stars, IAU Symposium No.\ 182, eds.\ B.\
Reipurth \& C.\ Bertout, (Dordrecht: Kluwer Academic Publishers), 93
\reference{} Eisl\"offel, J., Davis, C. J., Ray, T. P., \& Mundt, R. 1994,
\apj, 422, L91
\reference{} Eisl\"offel, J., Smith, M.D., Davis, C.J., \& Ray, T.P. 1996,
\aj, 112, 2087
\reference{} Eisl\"offel, J., \& Mundt, R. 1994, \aap, 284, 530
\reference{} Everett, M.E. 1997, \apj, 478, 246
\reference{} Franco, G.A.P. 1990, \aap, 227, 499
\reference{} Gredel, R. 1996, A\&A, 305, 582
\reference{} Heathcote, S, Morse, J.A., Hartigan, P., Reipurth, B.,
Schwartz, R.D., Bally, J., \& Stone, J.M. 1996, \aj, 112, 1141
\reference{} Mundt, R., \& Fried, J.W. 1983, \apj, 274, L83
\reference{} Noriega-Crespo, A., Garnavich, P.M., Curiel, S., Raga, A.C.,
\& Ayala, S. 1997, \apj, 486, L55
\reference{} Olberg, M., Reipurth, B., \& Booth, R.S. 1992, \aap,
259, 252
\reference{} Padman, R., Bence, S., \& Richer, J. 1997, in Herbig-Haro
Outflows and the Birth of Low Mass Stars, IAU Symposium No.\ 182, eds.\ B.\
Reipurth \& C.\ Bertout, (Dordrecht: Kluwer Academic Publishers), 123
\reference{} Reipurth, B., \& Heathcote, S., 1991 \aap, 246, 511
\reference{} Schwartz, R., Jones, B.F., \& Sirk, M. 1984, \aj, 89, 1735
\reference{} Smith, M.D., \& Brand, P.W.J.L., 1990, MNRAS, 242, 495
\reference{} Smith, M.D. 1991, MNRAS, 252, 378
\reference{} Smith, M.D. 1994, MNRAS, 266, 238
\reference{} Suttner, G., Smith, M.D., Yorke, H.W., \& Zinnecker, H. 1997,
A\&A,
318, 595
\reference{} Taylor, S.D., \& Raga, A.C., 1995 \aap, 296, 823
\reference{} Zinnecker, H., McCaughrean, M., \& Rayner, J. 1997, in Low Mass
Star Formation -- from Infall to Outflow, eds.\ F.\ Malbet \& A.\ Castets,
(Observatoire de Grenoble), 198
\end{references}
\end{document}